\documentclass[useAMS,showkeys]{tMPH2e}

\usepackage{amsmath}

\newcommand\beq{\begin{equation}}
\newcommand\eeq{\end{equation}}
\newcommand\beqa{\begin{eqnarray}}
\newcommand\eeqa{\end{eqnarray}}
\newcommand{\nn}{\nonumber\\}
\newcommand{\Ss}{\text{SS}}
\newcommand{\ps}{\text{PS}}
\newcommand{\hs}{\text{HS}}
\newcommand{\dd}{\text{d}}
\newcommand{\ee}{\text{e}}

\begin{document}

\title{Are the energy and virial routes  to thermodynamics equivalent for hard spheres?}

\author{Andr\'es Santos\thanks{$^\ast$
Email: andres@unex.es \vspace{10pt}
\newline\centerline{\tiny{ {\em Molecular Physics}}}
\newline\centerline{\tiny{ISSN 0026-8976 print/ ISSN
1362-3028
 online
\textcopyright 2005 Taylor \& Francis Ltd}}
\newline\centerline{\tiny{ http://www.tandf.co.uk/journals}}
\newline \centerline{\tiny{DOI:
10.1080/002689700xxxxxxxxxxxx}}}$^\ast$
 \\\vspace{6pt}
Departamento de F\'{\i}sica, Universidad de Extremadura, E-06071
Badajoz, Spain}

\received{\today}

\label{firstpage} \doi{10.1080/002689700xxxxxxxxxxx}

\issn{1362-3028}  \issnp{0026-8976} 

\markboth{A. Santos}{Are the energy and virial routes  to
thermodynamics equivalent for hard spheres?}

\maketitle

\begin{abstract}
The internal energy of hard spheres (HS) is the same as that of an
ideal gas, so that the energy route to thermodynamics becomes
useless. This problem can be avoided by taking an interaction
potential that reduces to the HS one in certain limits. In this
paper the square-shoulder (SS) potential characterized by a
hard-core diameter $\sigma'$, a soft-core diameter $\sigma>\sigma'$
and a shoulder height $\epsilon$ is considered. The SS potential
becomes the HS one if (i) $\epsilon\to 0$, or (ii)
$\epsilon\to\infty$, or (iii) $\sigma'\to\sigma$ or (iv) $\sigma'\to
0$ and  $\epsilon\to\infty$. The energy-route equation of state for
the HS fluid is obtained in terms of the radial distribution
function for the SS fluid by taking the limits (i) and (ii). This
equation of state is shown to exhibit, in general, an artificial
dependence on the diameter ratio $\sigma'/\sigma$. If furthermore
the limit $\sigma'/\sigma\to 1$ is taken, the resulting equation of
state for HS coincides with that obtained through the virial route.
The necessary and sufficient condition to get thermodynamic
consistency between both routes for arbitrary $\sigma'/\sigma$ is
derived.
\end{abstract}

\section{Introduction\label{sec1}}
As is well known, there exist several routes to obtain the
thermodynamic quantities of a fluid in equilibrium in terms of the
pair interaction potential $\phi(r)$ and the  radial distribution
function $g(r)$ \cite{B74,BH76,HM86}. The most frequently used are
the virial route,
\beq
\frac{\beta p}{\rho}\equiv Z=1+\frac{2\pi}{3}\rho\int_0^\infty \dd
r\, r^3 y(r)\frac{\partial}{\partial r} \ee^{-\beta \phi(r)},
\label{1}
\eeq
the compressibility route
\beq
\left(\beta\frac{\partial p}{\partial \rho}\right)^{-1}\equiv
\chi=1+4\pi\rho\int_0^\infty \dd r\, r^{2}\left[\ee^{-\beta
\phi(r)}y(r)-1\right],
\label{2}
\eeq
and the energy route
\beq
u=\frac{3}{2\beta}+2\pi\rho\int_0^\infty \dd r\,
r^{2}\phi(r)\ee^{-\beta \phi(r)} y(r).
\label{3}
\eeq
In Eqs.\ (\ref{1})--(\ref{3}), $p$ is the pressure, $\rho$ is the
number density, $\beta\equiv 1/k_BT$ is the inverse temperature, $Z$
is the compressibility factor,  $\chi$ is the (dimensionless)
isothermal compressibility, $u$ is the internal energy per particle
and $y(r)\equiv \exp[\beta\phi(r)]g(r)$ is the cavity (or
background) function.

The three thermodynamic quantities $Z$, $\chi$ and $u$ are connected
by thermodynamic relations, namely
\beq
\chi^{-1}=\frac{\partial}{\partial\rho}\left(\rho Z\right),
\label{4}
\eeq
\beq
\rho\frac{\partial}{\partial\rho}u=\frac{\partial}{\partial
\beta}{Z}.
\label{5}
\eeq
Thus, the compressibility factor, and hence the equation of state
(EoS), can be obtained from $y(r)$ [or, equivalently, $g(r)$] either
directly from Eq.\ (\ref{1}), or from Eqs.\ (\ref{2}) and (\ref{4}),
or from Eqs.\ (\ref{3}) and (\ref{5}). Given an interaction
potential $\phi(r)$, if the \textit{exact} cavity function $y(r)$ is
known for any thermodynamic state $(\rho,\beta)$, the three routes
yield of course the same  EoS. On the other hand, if an
\textit{approximate} function $y(r)$ is used, a different result is
in general obtained from each route, a problem known as
thermodynamic inconsistency of the approximation.

Some liquid state theories contain one or more adjustable
state-dependent parameters which are tuned to achieve thermodynamic
consistency between several routes, usually the compressibility and
the virial ones. Examples include, among other approaches,  the
modified hypernetted-chain closure \cite{RA79}, the Rogers--Young
closure \cite{RY84}, the Zerah--Hansen closure \cite{ZH85}, the
self-consistent Ornstein--Zernike approximation \cite{HS77}, the
hierarchical reference theory \cite{PR95}, Lee's theory based on the
zero-separation theorems \cite{L95}, the generalized mean spherical
approximation \cite{W73} or the rational-function approximation
\cite{YS91}.

On the other hand, standard  theories do not have fitting parameters
and thus they are in general thermodynamically inconsistent.  An
interesting result, however, is that the hypernetted-chain (HNC)
integral equation provides thermodynamically consistent results
through the virial and energy routes, regardless of the potential
$\phi(r)$ \cite{nBH76}. A similar result has recently been reported
\cite{MFKN05} in the case of the mean spherical approximation (MSA)
applied to soft potentials, such as the Gaussian core model
$\phi(r)=\epsilon\exp[-(r/\sigma)^2]$.  Therefore, a certain close
relationship between the energy and virial routes seems to exist, at
least for some approximate theories and/or some interaction models.
This has been further supported by a recent proof on the equivalence
between both routes when taking a hard-sphere (HS) limit from the
square-shoulder (SS) model, regardless of the approximate theory
used to describe the structural properties of the fluid \cite{S05}.

\begin{figure}
 \centerline{\epsfbox{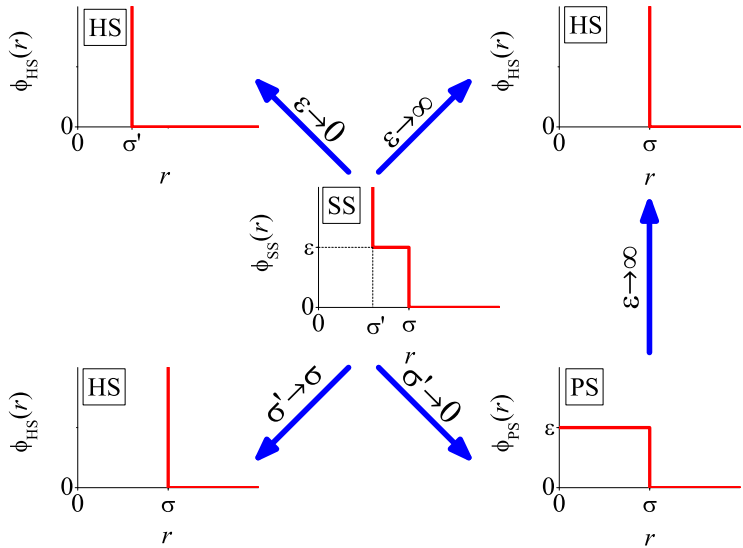}}
 \caption{(Color online) The graph in the center represents the square-shoulder (SS) interaction potential. It reduces
to the hard-sphere (HS) potential of hard-core diameter $\sigma'$ in
the limit $\epsilon\to 0$ and to the HS potential of diameter
$\sigma$ in the limit $\epsilon\to\infty$ as well as in the limit
$\sigma'\to\sigma$. The penetrable-sphere (PS) interaction model is
obtained in the limit $\sigma'\to 0$. The limit $\epsilon\to\infty$
taken in the PS potential leads again to the HS model.}
\label{fig1}
\end{figure}

In a HS liquid, the second term on the right-hand side of Eq.\
(\ref{3}) vanishes, so that the internal energy per particle is the
same as that of an ideal gas, i.e., it reduces to the kinetic
contribution $3/2\beta$ and is independent of density. Moreover, the
compressibility factor $Z$ of a HS liquid is independent of
temperature. As a consequence, the thermodynamic relation (\ref{5})
is trivially satisfied as $0=0$ and it is in principle impossible to
obtain the EoS of the HS fluid through the energy route. This ill
definition of the energy route  of a HS fluid can be avoided by
considering a suitable interaction potential that reduces to the HS
one in certain limits. The simplest choice for such a potential is
perhaps the SS function
\beq
\phi_\Ss(r)=\begin{cases} \infty,&r<\sigma',\\
\epsilon,&\sigma'<r<\sigma,\\
0,&r>\sigma,
\end{cases}
\label{9}
\eeq
where $\epsilon$ is a positive constant that measures the height of
the square shoulder, while the width is given by the difference
$\sigma-\sigma'$. The potential (\ref{9}) has been studied by
several authors in different contexts  \cite{SS}. For the purpose of
this paper, it is chosen here because it reduces to the HS
interaction potential in several special cases. First, in the limit
of zero shoulder height, $\epsilon\to 0$ (or, equivalently, in the
limit of infinite temperature), the potential (\ref{9}) becomes that
of HS of diameter $\sigma'$. In the opposite limit of infinite
shoulder height, $\epsilon\to\infty$ (or, equivalently, in the limit
of zero temperature), one also gets the HS interaction, but this
time the one corresponding to a diameter $\sigma$. These two limits
are important because one of them is needed as a boundary condition
when integrating the internal energy over temperature to get  $Z$
from Eq.\ (\ref{5}). Interestingly, the HS potential of diameter
$\sigma$ is also recovered  in the limit of zero shoulder width,
$\sigma'\to\sigma$, at finite $\epsilon$ (finite temperature). In
the opposite limit $\sigma'\to 0$, the hard-core part of the
interaction has a vanishing influence and the SS potential becomes
that of so-called penetrable spheres (PS). The PS fluid has been
extensively studied \cite{PS} as an example of bounded potentials
describing the effective two-body interaction in some colloidal
systems \cite{L01}. These special limits of the SS potential are
sketched in Fig.\ \ref{fig1}. Apart from that, it is worth recalling
that the square-well (SW) potential is described by Eq.\ (\ref{9}),
except that $\epsilon$ is negative in that case. {}From the SW
potential one can still get the HS potential of diameter $\sigma'$
in the limit $|\epsilon|\to 0$ (or, equivalently, in the limit
$T\to\infty$, what then provides the boundary condition for the
energy route to the EoS) and the HS potential of diameter $\sigma$
in the limit $\sigma'\to\sigma$, but not either of the other two
cases ($\sigma'\to 0$ and $|\epsilon|\to\infty$) depicted in Fig.\
\ref{fig1}.

As said before, starting from the EoS for SS particles obtained via
the energy route, it can be proven \cite{S05} that   when one
\textit{first} takes the limit $\epsilon\to \infty$ and
\textit{then} the limit $\sigma'\to\sigma$, the resulting EoS for HS
always coincides with the one obtained via the virial route.
However, there is no \textit{a priori} reason to expect that this
equivalence still holds when the HS limit is taken following a
different path. The aim of this paper is to clarify this issue and
show, by taking a couple of alternative paths, that the energy-route
EoS depends indeed on the path followed from SS to HS.

This paper is organized as follows. The virial and energy routes to
the thermodynamic properties of SS fluids are presented in section
\ref{sec2}. The HS limits along different paths are worked out in
section \ref{sec3}. The results are summarized and discussed in
section \ref{sec4}. Finally, the paper is closed with the conclusion
section.

\section{Virial and energy routes to the equation of state of square-shoulder fluids\label{sec2}}
Let us consider a fluid of particles interacting via the pairwise SS
potential          given by Eq.\ (\ref{9}). Therefore,
\beq
\ee^{-\beta\phi_\Ss(r)}=\begin{cases} 0,&r<\sigma',\\
\ee^{-\beta\epsilon},&\sigma'<r<\sigma,\\
1,&r>\sigma.
\end{cases}
\label{9bis}
\eeq

We take the ``outer'' diameter $\sigma$ as fixed and define the
relative value $\lambda\equiv \sigma'/\sigma$ of the ``inner''
diameter $\sigma'$. This quantity $0\leq\lambda\leq 1$ parameterizes
a family of independent SS potentials. Given a value of $\lambda$,
the thermodynamic state is determined by the density and the
temperature. In dimensionless units, we can measure the density by
the parameter $\eta\equiv (\pi/6)\rho\sigma^3$ \cite{note1} and the
temperature by $\beta^*\equiv \beta\epsilon=\epsilon/k_BT=1/T^*$.
Moreover, without loss of generality, henceforth the distances will
be understood to be measured in units of $\sigma$. For the SS
interaction (\ref{9}), the virial and energy equations, Eqs.\
(\ref{1}) and (\ref{3}), become
\beq
Z_{\text{SS}}^v(\eta,\beta^*;\lambda)=1+4\eta\left[\lambda^3
\ee^{-\beta^*}y_{\text{SS}}(r=\lambda|\eta,\beta^*;\lambda)+
(1-\ee^{-\beta^*})y_{\text{SS}}(r=1|\eta,\beta^*;\lambda)\right],
\label{2.1}
\eeq
\beq
u_{\text{SS}}(\eta,\beta^*;\lambda)=\epsilon\left[\frac{3}{2\beta^*}+12\eta
\ee^{-\beta^*}\int_{\lambda}^1 \dd r\, r^2
y_{\text{SS}}(r|\eta,\beta^*;\lambda)\right],
\label{2.3}
\eeq
respectively. The notation in Eqs.\ (\ref{2.1}) and (\ref{2.3})
makes explicit the dependence of the SS quantities on $\eta$,
$\beta^*$ and $\lambda$. Besides, the superscript $v$ in Eq.\
(\ref{2.1}) is introduced to emphasize that it corresponds to the
virial route. Making use of Eq.\ (\ref{5}), the energy equation
(\ref{2.3}) yields the following expression for the compressibility
factor,
\beq
 Z_{\text{SS}}^e(\eta,\beta^*;\lambda)=Z_{\text{HS}}^e(\eta\lambda^3)+12\eta\frac{\partial}{\partial\eta}\eta
 \int_0^{\beta^*} \dd \beta_1^*
 \,
\ee^{-\beta_1^*}\int_{\lambda}^1 \dd r\, r^2
y_{\text{SS}}(r|\eta,\beta_1^*;\lambda),
\label{2.4}
\eeq
where use has  already been made of the mapping SS$\to$HS in the
infinite-temperature limit $\beta^*\to 0$. The superscript $e$
affects to  $Z_{\text{SS}}$ as well as to  $Z_{\text{HS}}$ since
both sides of Eq.\ (\ref{2.4}) must agree in the limit $\beta^*\to
0$.

The virial series expansions for the cavity function and the
compressibility factor of the SS fluid are defined as
\beq
y_{\text{SS}}(r|\eta,\beta^*;\lambda)=1+\sum_{n=1}^\infty
y_n^{\text{SS}}(r|\beta;\lambda)\eta^{n},
\label{2.10}
\eeq
\beq
Z_{\text{SS}}(\eta,\beta^*;\lambda)=1+\sum_{n=1}^\infty
b_{n+1}^{\text{SS}}(\beta;\lambda)\eta^{n}.
\label{2.11}
\eeq
Inserting the expansion (\ref{2.11}) into Eqs.\ (\ref{2.1}) and
(\ref{2.4}), one gets the following expressions for the virial-route
and energy-route virial coefficients in the SS model:
\beq
b_n^{\Ss,v}(\beta^*;\lambda)=4\left[\lambda^3
\ee^{-\beta^*}y_{n-2}^{\text{SS}}(\lambda|\beta^*;\lambda)+
(1-\ee^{-\beta^*})y_{n-2}^{\text{SS}}(1|\beta^*;\lambda)\right],\quad
n\geq 2,
\label{2.1k}
\eeq
\beq
b_n^{\Ss,e}(\beta^*;\lambda)=b_n^{\hs,e}\lambda^{3(n-1)}+12(n-1)
\int_0^{\beta^*} \dd \beta_1^*
 \,
\ee^{-\beta_1^*}\int_{\lambda}^1 \dd r\, r^2
y_{n-2}^{\text{SS}}(r|\beta_1^*;\lambda),\quad n\geq 2.
\label{2.12}
\eeq

\section{Hard-sphere limits \label{sec3}}
Given an (approximate) cavity function
$y_{\text{SS}}(r|\eta,\beta_1^*;\lambda)$, the compressibility
factor obtained from Eq.\ (\ref{2.4}) differs in general from the
one given by Eq.\ (\ref{2.1}). The question is, does that difference
persist in the HS limit? To address this point, we need to specify
the path followed to get the HS limit.

\begin{figure}
 \centerline{\epsfxsize=7cm\epsfbox{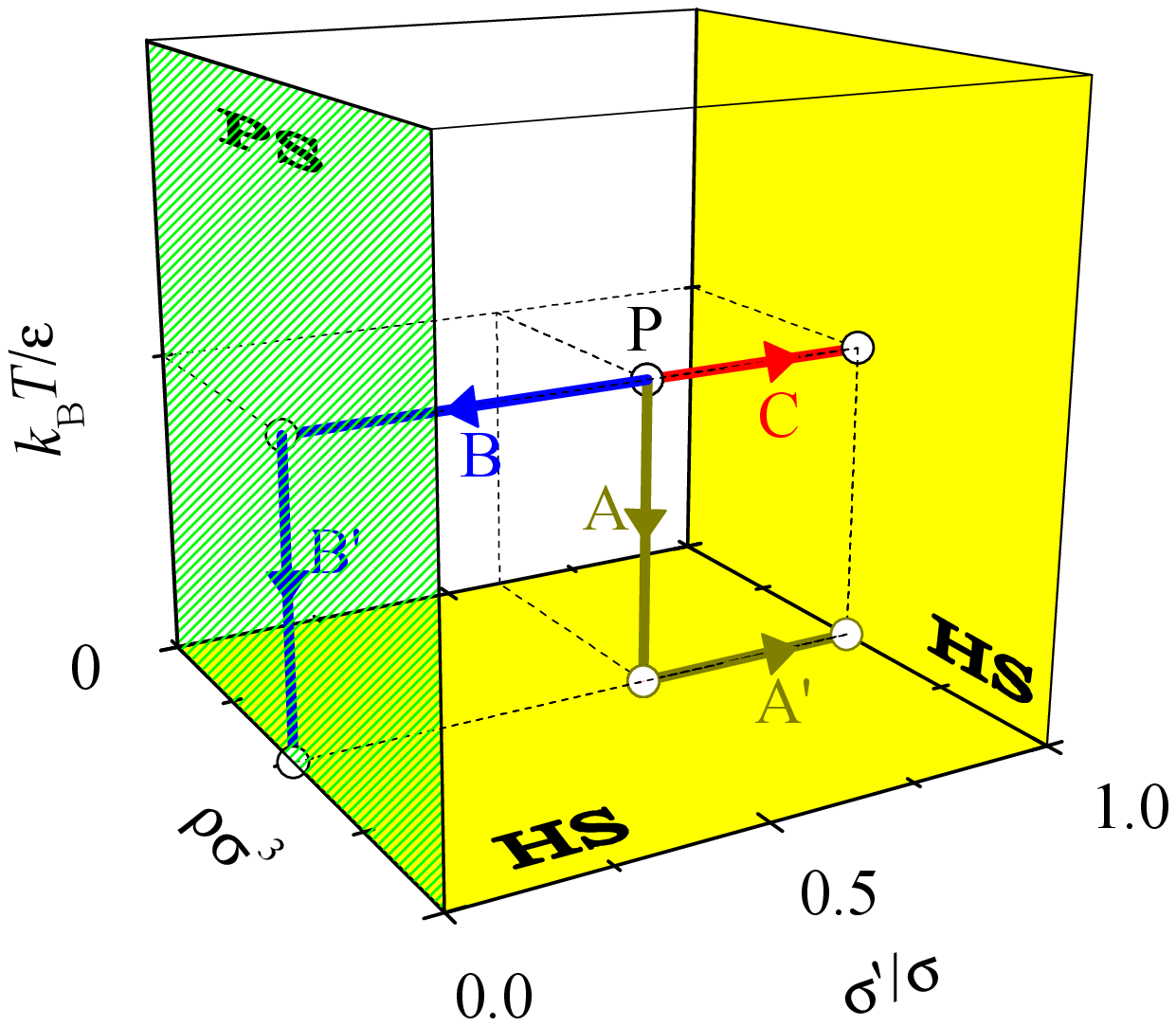}}
 \caption{(Color online) Parameter space for SS fluids. The plane
 $\sigma'/\sigma=0$ represents the PS fluids, while the planes
$k_BT/\epsilon=0$ and $\sigma'/\sigma=1$ correspond to  HS fluids.
Starting from  a given SS fluid (represented by the point P), it is
possible to go to the HS fluid at the same density by following
different paths. In particular, the paths A, A+A', B+B' and C are
considered in the paper.}
\label{fig2}
\end{figure}

Figure \ref{fig2} represents the relevant three-dimensional
parameter space for SS fluids. The plane $\sigma'/\sigma=\lambda=0$
represents the PS two-dimensional parameter space. The planes
$k_BT/\epsilon=1/\beta^*=0$ and $\sigma'/\sigma=\lambda=1$
correspond to HS of diameter $\sigma$, whose dimensionless
properties depend only on the reduced density
$\eta=(\pi/6)\rho\sigma^3$ and  should be independent of $\lambda$
(on the plane $k_BT/\epsilon=0$) and of temperature (on the plane
$\sigma'/\sigma=1$). The point P represents a  SS liquid with a
given value of $\lambda$ and at a given thermodynamic state
$(\eta,\beta^*$). {}From it, one can follow several paths by
changing $\lambda$ and/or $\beta^*$ to reach a HS liquid at the same
density $\eta$. Here we will be concerned with paths A (i.e.,
$\beta^*\to\infty$), A+A' ($\beta^*\to\infty$ followed by
$\lambda\to 1$), B+B' ($\lambda\to 0$ followed by
$\beta^*\to\infty$) and C ($\lambda\to 1$).

\subsection{Path A+A'}
Let us  take the zero temperature limit, $\beta^*\to\infty$ in both
sides of Eq.\ (\ref{2.4}). The result is
\beq
 Z_{\text{HS}}^e(\eta)-Z_{\text{HS}}^e(\eta\lambda^3)=12\eta\frac{\partial}{\partial\eta}\eta\int_0^{\infty} \dd \beta^*
 \,
\ee^{-\beta^*}\int_{\lambda}^1 \dd r\, r^2
y_{\text{SS}}(r|\eta,\beta^*;\lambda).
\label{2.5}
\eeq
This is actually the result obtained through path A, which will be
analyzed later on. Before proceeding with the subsequent step A',
let us rewrite Eq.\ (\ref{2.5}) in an equivalent form. First, note
that $Z_{\text{HS}}(\eta\lambda^3)$ can be expanded around $\eta$:
\beq
Z_{\text{HS}}(\eta\lambda^3)=Z_{\text{HS}}(\eta)+\sum_{n=1}^\infty
\frac{(-\eta)^n (1-\lambda^3)^n}{n!}\frac{\partial^n}{\partial
\eta^n} Z_{\text{HS}}(\eta).
\label{2.6}
\eeq
As a consequence, Eq.\ (\ref{2.5}) yields
\beq
Z_{\text{HS}}^e(\eta)+\sum_{n=1}^\infty \frac{(-\eta)^n
(1-\lambda^3)^n}{(n+1)!}\frac{\partial^n}{\partial \eta^n}
Z_{\text{HS}}^e(\eta)=1+\frac{12\eta}{1-\lambda^3} \int_0^{\infty}
\dd \beta^*
 \,
\ee^{-\beta^*}\int_{\lambda}^1 \dd r\, r^2
y_{\text{SS}}(r|\eta,\beta^*;\lambda),
\label{2.7}
\eeq
where use has been made of the consistency condition
$Z_{\text{HS}}(0)=1$. Taking now the limit $\lambda\to 1$ (path A')
in both sides of Eq.\ (\ref{2.7}), and using the physical condition
\beq
\lim_{\lambda\to 1}
y_{\text{SS}}(r|\eta,\beta^*;\lambda)=y_{\text{HS}}(r|\eta),
\label{2.8}
\eeq
we finally get
\beq
Z_{\text{HS}}^e(\eta)=1+4\eta y_{\text{HS}}(1|\eta).
\label{2.9}
\eeq
This is not but the virial EoS, Eq.\ (\ref{2.1}), particularized to
HS. This proves that, no matter which approximation is used to get
$y_{\text{SS}}(r|\eta,\beta^*;\lambda)$, the energy EoS coincides
with the virial one when the HS limit is reached from the SS fluid
following the double path A+A'. The proof presented in Ref.\
\cite{S05} is slightly more general since it applies to mixtures and
to any dimensionality.

The equivalence between the energy an virial routes when  the path
A+A' is followed can also be proven at the level of the virial
coefficients. We first take the limit $\beta^*\to \infty$ on both
sides of Eq.\ (\ref{2.12}) (path A) with the result
\beq
b_n^{\hs,e}=\frac{12(n-1)}{1-\lambda^{3(n-1)}} \int_0^{\infty} \dd
\beta^*
 \,
\ee^{-\beta^*}\int_{\lambda}^1 \dd r\, r^2
y_{n-2}^{\text{SS}}(r|\beta^*;\lambda),\quad n\geq 2.
\label{3.1}
\eeq
Next, the limit $\lambda\to 1$ (path A') yields
\beq
b_n^{\hs,e}=b_n^{\hs,v}=4 y_{n-2}^{\text{HS}}(1),\quad n\geq 2.
\label{3.2}
\eeq

\subsection{Path B+B'}
\begin{figure}
 \centerline{\epsfbox{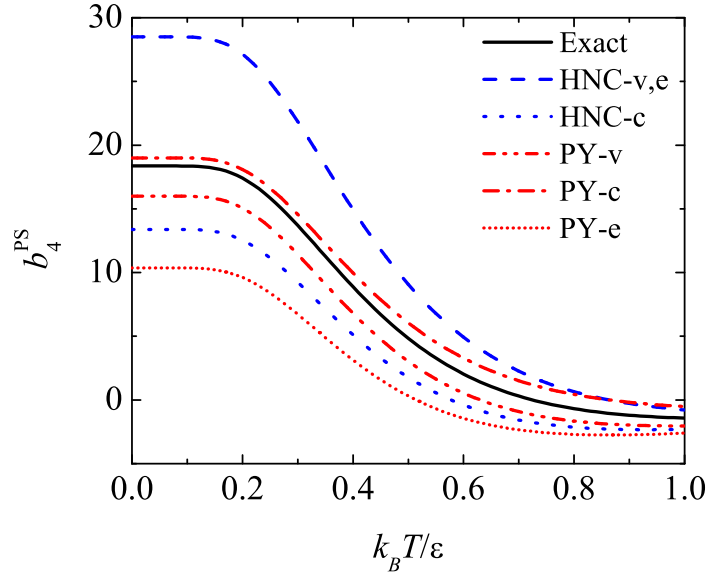}}
\caption{(Color online)  Temperature dependence of the fourth virial
coefficient for the PS fluid. The curves correspond to the exact
result    (---),    the virial and energy routes in the HNC
approximation    (-- -- --),    the compressibility route in the HNC
approximation    (- - -),    the virial route in the PY
approximation    (-- $\cdot$ $\cdot$ --),  the compressibility route
in the PY approximation    (-- $\cdot$ -- $\cdot$)     and the
energy route in the PY approximation    ($\cdots$).}
\label{fig3}
\end{figure}

Let us consider now the path B+B' in Fig.\ \ref{fig2}, which is very
different from the path A+A' considered above. We will restrict
ourselves to the fourth virial coefficient since this is enough to
check that the virial and energy routes do not coincide now in the
HS limit. We start taking the PS limit $\lambda\to 0$ (path B) on
both sides of Eqs.\ (\ref{2.1k}) and (\ref{2.12}),
\beq
b_n^{\ps,v}(\beta^*)=4\left(1-\ee^{-\beta^*}\right)
y_{n-2}^{\text{PS}}(1|\beta_1^*),\quad n\geq 2,
\label{3.3}
\eeq
\beq
b_n^{\ps,e}(\beta^*)=12(n-1) \int_0^{\beta^*} \dd \beta_1^*
 \,
\ee^{-\beta_1^*}\int_{0}^1 \dd r\, r^2
y_{n-2}^{\text{PS}}(r|\beta_1^*),\quad n\geq 2.
\label{3.4}
\eeq
The exact function $y_{2}^{\text{PS}}(r|\beta^*)$ for the PS model,
as well as the corresponding expressions in the Percus--Yevick  (PY)
and HNC approximations, have recently been obtained \cite{SM06}. In
particular, the PY result is
\beqa
y_{2}^{\text{PS-PY}}(r|\beta^*)&=&\frac{\left(1-\ee^{-\beta^*}\right)^3}{35
r}\left[(r-1)^4(r^3+4r^2-53r-162)\left(4
\ee^{-\beta^*}-1\right)\Theta(1-r)\right.\nn
&&+2(r-2)^2(r^5+4r^4-51r^3-10r^2+479r-81)\left(1-\ee^{-\beta^*}\right)
\Theta(r-2)\nn &&\left.-(r-3)^4(r^3+12r^2+27r-6)\Theta(3-r)\right],
\label{3.5}
\eeqa
where $\Theta(x)$ is the Heaviside step function. Inserting Eq.\
(\ref{3.5}) into Eqs.\ (\ref{3.3}) and (\ref{3.4}), one gets
\beq
b_4^{\text{PS-PY},v}(\beta^*)=\frac{16}{35}\left(1-\ee^{-\beta^*}\right)^4\left(35-171\ee^{-\beta^*}\right),
\label{3.6}
\eeq
\beq
b_4^{\text{PS-PY},e}(\beta^*)=\frac{2}{175}\left(1-\ee^{-\beta^*}\right)^4\left(907-6347\ee^{-\beta^*}\right),
\label{3.7}
\eeq
respectively. As expected, the virial and energy routes to the
fourth virial coefficient of the PS fluid differ in the PY
approximation. This difference persists in the HS zero-temperature
limit $\beta^*\to\infty$ (path B'), namely,
\beq
b_4^{\text{HS-PY},v}=16,\quad
b_4^{\text{HS-PY},e}=\frac{1814}{175}\simeq 10.37.
\label{3.8}
\eeq
However,
$b_4^{\text{HNC-PY},v}(\beta^*)=b_4^{\text{HNC-PY},e}(\beta^*)$
\cite{SM06}, in agreement with a general property of the HNC
approximation \cite{nBH76}. Figure \ref{fig3} shows the temperature
dependence of the exact fourth virial coefficient of the PS fluid,
as well as the results obtained from the PY and HNC approximations
via the virial, compressibility and energy routes \cite{SM06}.

\subsection{Path A}
Equations\ (\ref{2.9}) and (\ref{3.2}) show that the virial and
energy routes are always equivalent when the HS limit is taken
through the double path A+A'. On the other hand, Eq.\ (\ref{3.8})
shows that this equivalence is generally broken when the chosen path
is B+B'. While in the path A+A' one ends with $\lambda=1$, in the
path B+B' the first step is $\lambda\to 0$. Therefore, it might be
reasonably expected that if one directly goes from SS to HS through
the path A (see Fig.\ \ref{fig2}) the resulting energy-route EoS for
HS \textit{artificially} depends on $\lambda$. To illustrate this,
let us consider the following toy approximation for the function
$y_2^\Ss (r|\beta^*;\lambda)$:
\beq
y_2^{\text{SS-toy}}(r|\beta^*;\lambda)=y_2^{\text{PS-PY}}(r|\beta^*)+y_2^{\text{HS-PY}}(r/\lambda)\lambda^6-
\frac{y_2^{\text{PS-PY}}(r|\beta^*)y_2^{\text{HS-PY}}(r/\lambda)\lambda^6}{y_2^{\text{HS-PY}}(r)},
\label{3.9}
\eeq
where $y_2^{\text{PS-PY}}(r|\beta^*)$ is given by Eq.\ (\ref{3.5})
and
$y_2^{\text{HS-PY}}(r)=\lim_{\beta^*\to\infty}y_2^{\text{PS-PY}}(r|\beta^*)$.
The toy approximation (\ref{3.9}) reduces to the PY results in the
four limits indicated in Fig.\ (\ref{fig1}), namely
\beq
\lim_{\beta^*\to
0}y_2^{\text{SS-toy}}(r|\beta^*;\lambda)=y_2^{\text{HS-PY}}(r/\lambda)\lambda^6,
\label{3.10}
\eeq
\beq
\lim_{\beta^*\to
\infty}y_2^{\text{SS-toy}}(r|\beta^*;\lambda)=\lim_{\lambda\to
1}y_2^{\text{SS-toy}}(r|\beta^*;\lambda)=y_2^{\text{HS-PY}}(r),
\label{3.11}
\eeq
\beq
\lim_{\lambda\to
0}y_2^{\text{SS-toy}}(r|\beta^*;\lambda)=y_2^{\text{PS-PY}}(r|\beta^*).
\label{3.12}
\eeq
{}From that point of view, $y_2^{\text{SS-toy}}(r|\beta^*;\lambda)$
can be seen as a simplified version of the true
$y_2^{\text{SS-PY}}(r|\beta^*;\lambda)$ provided by the PY
approximation. In any case, the point here is not how accurate or
how close to the PY function the toy approximation (\ref{3.9}) is,
but to illustrate the sensitivity of the energy route to the fixed
value of $\lambda$.

\begin{figure}
 \centerline{\epsfbox{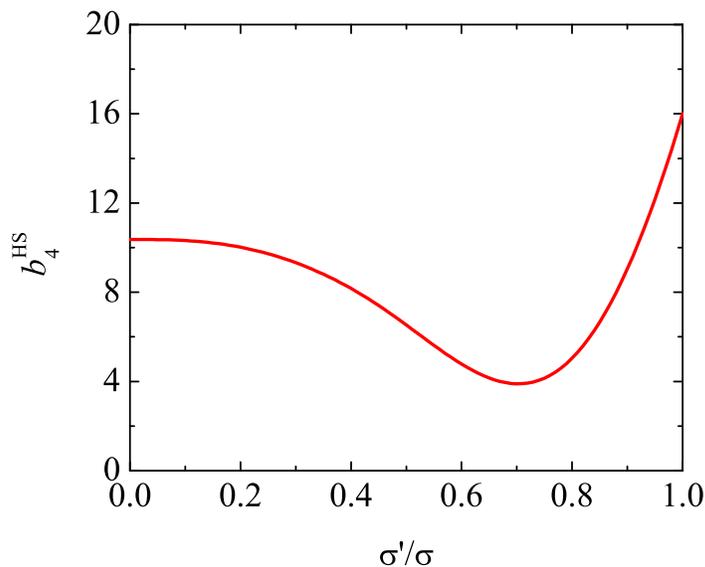}}
\caption{(Color online)  Plot of the HS fourth virial coefficient as
a function of $\lambda\equiv\sigma'/\sigma$. This coefficient is
obtained by starting from the energy-route coefficient for the SS
fluid in the approximation (\protect\ref{3.9}) and then taking the
zero-temperature limit (path A).}
\label{fig4}
\end{figure}

According to Eq.\ (\ref{3.1}), the energy-route fourth virial
coefficient when the path A is followed becomes
\beq
b_4^{\text{HS-toy},e}=\frac{36}{1-\lambda^{9}} \int_{\lambda}^1 \dd
r\, r^2 \int_0^{\infty} \dd \beta^*
 \,
\ee^{-\beta^*}y_{2}^{\text{SS-toy}}(r|\beta^*;\lambda).
\label{3.13}
\eeq
The integration over $\beta^*$ is straightforward, while in the
integration over $r$ one needs to distinguish the cases $0\leq
\lambda\leq \frac{1}{3}$, $\frac{1}{3}\leq \lambda\leq\frac{1}{2}$
and $\frac{1}{2}\leq \lambda\leq 1$. The artificial dependence of
$b_4^{\text{HS-toy},e}$ on $\lambda$ is shown in Fig.\ \ref{fig4}.
The extreme points of the curve agree with the results previously
obtained, i.e., $b_4^{\text{HS-PY},e}=b_4^{\text{HS-PY},v}=16$ along
the path A+A' but $b_4^{\text{HS-PY},e}=1814/175\neq
b_4^{\text{HS-PY},v}$ along the path B+B'. It cannot be ascertained
at this point whether the non-monotonic behavior in Fig.\ \ref{fig4}
is an artifact of the toy approximation (\ref{3.9}) or is a feature
also shared by the PY approximation. In any case, the important
issue here is that the equivalence between the energy and virial
routes is only reached, in general, if $\lambda\to 1$.

\subsection{Path C}
As Fig.\ \ref{fig2} illustrates, the HS fluid can also be reached
from the PS fluid by keeping the temperature constant but shrinking
the shoulder (path C). However, this does not provide any
information about the energy-route EoS of the HS system. Taking the
limit $\lambda\to 1$ on both sides of Eq.\ (\ref{2.4}), one simply
gets $Z_\hs^e(\eta)=Z_\hs^e(\eta)$. In any case, an interesting
equation is obtained by first differentiating with respect to
$\lambda$ and then taking the limit $\lambda\to 1$. The result is
\beq
\left.\frac{\partial Z_\Ss^e(\eta,\beta^*;\lambda)}{\partial
\lambda}\right|_{\lambda=1}=3\eta\frac{\partial}{\partial\eta}\left[Z_\hs^e(\eta)-\left(1-\ee^{-\beta^*}\right)Z_\hs^v(\eta)\right].
\label{3.14}
\eeq

\section{Summary and discussion\label{sec4}}
The aim of this paper has been to investigate the possibility of
circumventing the ill definition of the energy route to
thermodynamics for HS fluids by first considering SS fluids and then
taking the adequate limits. The SS interaction potential is
particularly appropriate because it is simple and yet reduces to the
HS model in several independent limits (see Fig.\ 1). The
high-temperature limit ($T^*\to\infty$ or, equivalently,
$\epsilon\to 0$) is important since it provides the necessary
boundary condition to get the compressibility factor by integrating
the internal energy over temperature [see Eqs.\ (\ref{5}),
(\ref{2.4}) and (\ref{2.12})].

Imagine that an \textit{approximate} cavity function $y_\Ss
(r|\eta,\beta^*;\lambda)$ for the SS liquid is known (either
analytically or numerically). Then, Eqs.\ (\ref{2.4}) and/or
(\ref{2.12}) can be used to assign a meaning to the energy-route
compressibility factor for HS, $Z_\hs^e(\eta)$, or to the associated
virial coefficients, $b_n^{\hs,e}$. In order to do so, the limit of
vanishing shoulder width at finite temperature (path C in Fig.\
\ref{fig2}) is useless. On the other hand,  well defined results are
obtained by taking the zero-temperature limit $\beta^*\to\infty$
(path A), as shown by Eqs.\  (\ref{2.5}) and (\ref{3.1}). Although
the above limit is taken at fixed width $\lambda$, the final result
should, on physical grounds,  be independent of $\lambda$. However,
the approximate nature of $y_\Ss (r|\eta,\beta^*;\lambda)$ gives
rise, in general, to an inconsistent dependence of $Z_\hs^e(\eta)$
on $\lambda$, as illustrated by Fig.\ \ref{fig4} in the case of the
toy approximation (\ref{3.9}).

The artificial dependence of $Z_\hs^e(\eta)$ on $\lambda$ suggests
to   further take the limit $\lambda\to 1$ as the most ``sensible''
path to go from SS to HS (path A+A'). In this way, the HS fluid is
 reached twice, first by decreasing the temperature (or,
equivalently, increasing the shoulder height, $\epsilon\to\infty$)
and then by shrinking the shoulder width. As proven here and in
Ref.\ \cite{S05}, the resulting EoS coincides exactly with the one
obtained directly from the virial route, i.e.
$Z_\hs^e(\eta)=Z_\hs^v(\eta)$, no matter which approximate theory is
used.

As said above, the anomalous $\lambda$-dependence of $Z_\hs^e(\eta)$
when the path A is followed has been illustrated by considering a
simple toy approximation, Eq.\ (\ref{3.9}), and thus it might be
conjectured that such a dependence disappears when a most
``respectful'' theory is taken into account. However, this is not
the case, at least for the PY theory. When the zero-temperature
limit $\beta^*\to\infty$ is taken at $\lambda=0$ (i.e., from the PS
model, path B+B'), the resulting value of the fourth virial
coefficient ($b_4^{\text{HS-PY},e}\simeq 10.37$) strongly differs
from the value obtained in the opposite limit $\lambda\to 1$
($b_4^{\text{HS-PY},e}=b_4^{\text{HS-PY},v}=16$).

Of course, all the routes to thermodynamics self-consistently agree
if the \textit{exact} function $y_\Ss (r|\eta,\beta^*;\lambda)$ is
considered. Moreover, the HNC theory is known to yield consistent
thermodynamic properties via the energy and virial routes, for any
interaction potential \cite{nBH76}. This is  explicitly verified in
the case of the fourth virial coefficient for PS fluids \cite{SM06},
as shown in Fig.\ \ref{fig3}. Therefore, a plot (similar to that of
Fig.\ \ref{fig4})  of the fourth virial coefficient obtained
 from path A would show the constant value
$b_4^{\text{HS-HNC},e}=b_4^{\text{HS-HNC},v}=28.5$. The interesting
question is, which is the necessary and sufficient condition to get
thermodynamic properties independent of $\lambda$ when going from SS
to HS through path A? To address this question, define the quantity
\beq
\Phi_\Ss(\eta;\lambda)\equiv 12\eta \int_0^{\infty} \dd \beta^*
 \,
\ee^{-\beta^*}\int_{\lambda}^1 \dd r\, r^2
y_{\text{SS}}(r|\eta,\beta^*;\lambda).
\label{4.1}
\eeq
Then, either from Eq.\ (\ref{2.5}) or from Eq.\ (\ref{2.3}) using
the thermodynamic relation $u=\partial (\beta f)/\partial \beta$,
where $f$ is the free energy per particle, one gets
\beq
\Phi_\Ss(\eta;\lambda)=\varphi_\hs^e(\eta)-\varphi_\hs^e(\eta
\lambda^3),
\label{4.2}
\eeq
where $\varphi\equiv \beta f_{\text{ex}}$,  $f_{\text{ex}}$ being
the excess free energy per particle. Therefore, the sought necessary
and sufficient condition  is that the quantity
$\Phi_\Ss(\eta;\lambda)$ must be equal to a function that only
depends on $\eta$ plus a function that only depends on
$\eta'\equiv\eta\lambda^3$. The fact that both functions are
actually the same, except for a sign, is a consequence of the
trivial property $\Phi_\Ss(\eta;1)=0$. In differential form, the
condition (\ref{4.2}) becomes
\beq
\frac{\partial}{\partial \lambda}\lambda \frac{\partial}{\partial
\lambda}\Phi_\Ss(\eta;\lambda)=3\eta \frac{\partial}{\partial \eta}
\frac{\partial}{\partial \lambda}\Phi_\Ss(\eta;\lambda).
\label{4.3}
\eeq
If this condition is fulfilled, then Eq.\ (\ref{4.2}) implies that
$\varphi_\hs^e(\eta)=\Phi_\Ss(\eta;0)$. The independence of
$\varphi_\hs^e(\eta)$ on $\lambda$ implies that the energy  and
virial routes become equivalent, as discussed in the text. This can
be easily checked from Eq.\ (\ref{4.2}) by differentiating both
sides with respect to $\lambda$ and then setting $\lambda=1$. Taking
into account the thermodynamic relation $Z=\rho \partial (\beta
f)/\partial \rho$, one then gets
$Z_\hs^e(\eta)-1=Z_\hs^v(\eta)-1=4\eta y_\hs(1|\eta)$.

\section{Conclusion\label{sec5}}
The question posed in the title of the paper is  only meaningful if
the energy route is understood by starting from an interaction
potential $\phi(r)$ which encompasses the HS model in certain
limits. Since, according to Eq.\ \eqref{5}, the obtention of the
compressibility factor from the internal energy requires an
integration over temperature, it is necessary that the potential
$\phi(r)$ becomes  equivalent to that of HS (or negligible) in the
limit $T\to\infty$. Next, in order to get a non trivial result,
$\phi(r)$ must also become indistinguishable from HS in an
independent limit, for instance  $T\to 0$. The results presented in
this paper by taking $\phi(r)=\phi_{\text{SS}}(r)$ suggest that, in
general, either the energy-route EoS for HS fluids is identical to
the virial-route EoS or the former is not unique but depends on the
path followed to reach the HS fluid.

A final comment is in order. When considering the energy route in
the case of SW fluids, it is usual to fix the boundary condition at
$T\to\infty$ by freely choosing a convenient form for $Z_\hs(\eta)$,
such as the Carnahan--Starling EoS \cite{HM86}. However, this must
be done with caution since the resulting EoS would become
inconsistent when making the change
$\epsilon\leftrightarrow-\epsilon$, in which case
$\phi_{\text{SW}}(r)\leftrightarrow \phi_{\text{SS}}(r)$. {}From
that point of view, it would be more consistent to take the virial
form $Z_\hs^v(\eta)$ corresponding to the approximation (PY, HNC,
MSA, \ldots) being used to describe the SW fluid.

\bigskip

 This work has been supported by the Ministerio de Educaci\'on y
Ciencia (Spain) through Grant No. FIS2004-01399 (partially financed
by FEDER funds) and by the European Community's Human Potential
Programme under contract HPRN-CT-2002-00307, DYGLAGEMEM.


\begin{thebibliography}{99}
\bibitem{note1}
 Note that, in general, $\eta$ does not represent a packing
fraction. The true packing fraction is
$\eta'=(\pi/6)\rho{\sigma'}^3=\eta\lambda^3$. However, since
$\lambda$ will be considered as a varying parameter, it is
convenient to characterize the density with $\eta$ rather than with
$\eta'$.
\end{thebibliography}
\end{document}